
\magnification=\magstep1
\hsize=5in \vsize=7.8in
\baselineskip=13pt
\font\srm=cmr8
\font\sit=cmti8

\font\mrm=cmr9
\font\mit=cmti9
\font\mbf=cmbx9

\def\doublespace{\par\null\par}

\newcount\currentnumber
\def\currentno{\global\advance\currentnumber by 1
\the\currentnumber}

\def\heading#1{\doublespace\goodbreak
\noindent{\bf\currentno.~#1}\par}

\def\eqn#1{(\the\currentnumber.#1)}

\def\deg{{\rm deg}}
\def\gh{{\rm gh}}
\def\half{{1\over2}}
\def\Tr{{\rm Tr}}

\rightline{NI 92015}
\vfill
\centerline{\bf
A PATH-INTEGRAL APPROACH TO POLYNOMIAL
}\centerline{\bf
INVARIANTS OF LINKS%
\footnote{$^\ast$}{\srm
Condensed title: {\sit
A path-integral approach to polynomial invariants}, PACS numbers:
\goodbreak\noindent\parskip0pt
11.15.Tk, 02.40.+m, 03.65.Db.
}}
\doublespace
\centerline{
BOGUS\L AW BRODA%
\footnote{$^\dagger$}{\srm
Permanent address:
\sit
Department of Theoretical Physics, University of \L\'od\'z,
Pomorska 149/153, PL--90-236 \L\'od\'z, Poland.
}}
{\it\centerline{
Isaac Newton Institute for Mathematical Sciences}
\centerline{University of Cambridge, 20 Clarkson Road}
\centerline{Cambridge, CB3 0EH, UK}
}
{\rm\centerline{and}}
{\it\centerline{
Arnold Sommerfeld Institute for Mathematical Physics}
\centerline{Technical University of Clausthal, Leibnizstra\ss e 10}
\centerline{D-W--3392 Clausthal-Zellerfeld, FRG}
}
\vfill
{\leftskip1cm \rightskip1cm
\noindent
A brief review of a self-contained genuinely
three-dimensional mono\-d\-ro\-my-matrix based non-perturbative
covariant path-integral approach to {\it polynomial
invariants} of knots and links in the framework of
(topological) quantum Chern-Simons field theory is given.
An idea of ``physical'' observables represented by an
auxiliary topological quantum-mechanics model in an
external gauge field is introduced substituting rather a
limited notion of the Wilson loop.  Thus, the possibility
of using various generalizations of the Chern-Simons action
(also higher-dimensional ones) as well as a purely
functional language becomes open. The theory is quantized
in the framework of the best suited in this case {\it
antibracket-antifield} formalism of Batalin and Vilkovisky.
Using the Stokes theorem and formal translational
invariance of the path-integral measure a {\it monodromy
matrix} corresponding to an arbitrary pair of irreducible
representations of an arbitrary semi-simple Lie group is
derived.
\smallskip}
\vfill
\centerline{Preprint of the Newton Institute}
\vfill
\centerline{DECEMBER 1992}
\vfill\eject

\heading{
Introduction
}
Topological quantum field theory$^1$ (TQFT) has recently
become a fascinating and fashionable subject in
mathematical physics. Interestingly, the main source of
applications of TQFT is coming from mathematics (topology
of low-dimensional manifolds) rather than from physics
itself. The issue of classification of knots and links is
one of the most interesting ones in
low-dimensional topology.  To approach this problem one
usually tries to encode topology of a knot/link into some
algebraic structure. As was firstly indicated by
Witten,$^2$ the problem can be attacked by means of some
standard theoretical physics techniques of quantum
field theory. In particular, in the framework of
three-dimensional TQFT (Chern-Simons theory) not only can
all well-known polynomial invariants of knots and links be
derived but a lot of their generalizations as well.

Most of authors working on TQFT description of polynomial
invariants follow the original Witten's approach, which
heavily bases upon the underlying conformal field theory.
There is also a genuinely three-dimensional covariant
approach advocated in its perturbative version in Ref.~3. A
non-perturbative ``Hamiltonian'' version has been proposed
in Ref.~4.  A self-contained genuinely three-dimensional
non-perturbative covariant path-integral approach has been
firstly introduced in Ref.~5. The aim of our brief review
is to give a concise account of the developed form of the proposed idea, taking
as an example the simplest Chern-Simons model.

One should emphasize that the proposed approach has a number of advantages: It
is
self-contained, i.e. no notions of conformal field theory are
used explicitly or implicitly; it is genuinely
three-dimensional, i.e. there is no ``$2+1$''
decomposition; the approach is not limited to the
Chern-Simons description,$^6$ and it can easily be extended to
dimensions greater than three.$^7$

In Sect.~2 we introduce the classical Chern-Simons action,
which is next quantized in Sect.~3. In Sect.~4, using the
Stokes theorem, we derive the monodromy matrix, whereas
skein relations are dealt with in Sect.~5. Finishing
remarks (Sect.~6) concern a relation of the proposed approach to the
(quasi-)quantum group one.

\heading{
Classical Action
}
To begin with, we introduce, for an arbitrary semi-simple compact Lie
group $G$, the classical topological Chern-Simons action$^2$
on the three-dimensional sphere ${\cal S}^3$
$$
S_{\rm CS}={k\over4\pi}\int_{{\cal
S}^3}\Tr\left(A\wedge dA+{2\over3}A\wedge A\wedge
A\right)\qquad
$$
$$
\qquad={k\over4\pi}\int_{{\cal S}^3}d^3x\,
\varepsilon^{\mu\nu\lambda}\Tr\left(A_\mu\partial_\nu
A_\lambda+{2\over3}A_\mu A_\nu A_\lambda\right),
\qquad
\mu,\nu,\lambda=1,2,3,
\eqno\eqn{1}
$$
where $A_\mu=A_\mu^a(x)t_{\rm F}^a$ is the gauge potential,
valued in the fundamental representation $R_F$ of the Lie algebra $\cal G$ (the
Lie algebra of the
Lie group $G$) with standardly normalized antihermitian
generators, $\Tr\left(t_{\rm F}^at_{\rm
F}^b\right)=-\half\delta^{ab}$, and $k\in{\bf Z}^\pm$. For
any irreducible representation $R_i$ of $\cal G$ we have
$\left[t_i^a,t_i^b\right]=f^{abc}t_i^c$.
The use of \eqn{1} does not seem to be obligatory. One
could as well pick out the action of the, so-called,
BF-theory$^8$
$$
S_{\rm BF}={k\over4\pi}\int_{{\cal S}^3}d^3x\,
\varepsilon^{\mu\nu\lambda}\Tr(B_\mu F_{\nu\lambda}),
\eqno\eqn{2}
$$
where $B_\mu=B_\mu^a(x)t_{\rm F}^a$ is an auxiliary gauge
field, $F_{\mu\nu}$ is the strength of the gauge field, and
$k\in{\bf R}^\pm$. It appears that some generalization of
\eqn{2} containing the term $B^3$ is related to the square
of the modulus of \eqn{1}.$^{9}$ The action \eqn{2}
possesses some advantages in comparison with \eqn{1}: no
longer need $k$ be integer, and it can be easily
generalized to higher dimensions.$^7$ The action \eqn{1}
itself can also be generalized to higher dimensions (the inhomogeneous
case),$^{10}$ but for the sake of simplicity we will confine
ourselves to the standard homogeneous version.

\heading{
BV-Quantization
}
It appears that the most suitable for our purposes method
of quantization of gauge systems is the general {\it
antibracket-antifield} technique of Batalin and Vilkovisky
(BV).$^{11}$ In spite of the fact that the gauge symmetry
of Chern-Simons theory is {\it irreducible}, and obviously one could use the
standard BRST method, from our point of view, the BV-quantized
action is easier, for some technical reasons, to deal with.

In the framework of the BV approach one should find a
proper non-degenerated solution of the {\it master
equation}
$$
(S,S)=0,
\eqno\eqn{1}
$$
where ``$(\cdot,\cdot)$'' denotes the ({\it anti-})bracket in the
``extended phase space''.  Interestingly, the so-called,
{\it minimal part} of the solution of \eqn{1} can be put in
the following compact form, resembling the classical action
(2.1),
$$
S_{\rm min}={k\over4\pi}\int_{{\cal S}^3}\Tr\left({\cal
A}\wedge
d{\cal A}+{2\over3}{\cal A}\wedge{\cal A}\wedge{\cal
A}\right),
\eqno\eqn{2}
$$
where the {\it inhomogeneous field} ${{\cal A=A}^a(x)t_F^a}$
is defined by the formal sum of the fields  (gauge potential,
ghost) and their {\it antifields} (denoted with ``$\ast$'')
$$
{\cal A}=C+A+A^\ast+C^\ast.
\eqno\eqn{3}
$$
The form degrees and the ghost numbers of the components
entering $\cal A$ are as follows:
$$
\deg C=0,\qquad\gh C=1,
$$
$$
\deg A=1,\qquad\gh A=0,
$$
$$
\deg A^\ast=2,\qquad\gh A^\ast=-1,
$$
$$
\deg C^\ast=3,\qquad\gh C^\ast=-2.
\eqno\eqn{4}
$$
Taking into account the fact that only zero-ghost number
three-forms survive in the integrand of Eq.~\eqn{2}, we can
rewrite it explicitly as
$$
S_{\rm min}=S_{\rm CS}+{k\over4\pi}\int_{{\cal S}^3}d^3x\,
\varepsilon^{\mu\nu\lambda}\Tr\left(A_{\mu\nu}^\ast D_\lambda
C+\half C_{\mu\nu\lambda}^\ast C^2\right),
\eqno\eqn{5}
$$
where
$$
(D_\mu C)^a=\partial_\mu C^a+f^{abc}A_\mu^bC^c.
\eqno\eqn{6}
$$
The {\it auxiliary part} of the quantum action has the following
standard form
$$
S_{\rm aux}={1\over6}\int_{{\cal
S}^3}d^3x\,\varepsilon^{\mu\nu\lambda}\Tr\left(\bar C^\ast
B_{\mu\nu\lambda}\right),
\eqno\eqn{7}
$$
where $\bar C^\ast$ is the antifield antighost
corresponding to the antighost $\bar C$, and $B$ is the
Lagrange multiplier field. The form degrees and the ghost numbers are as
follows:
$$
\deg\bar C^\ast=0, \qquad \gh\bar C^\ast=0,
$$
$$
\deg B=3, \qquad \gh B=0,
$$
$$
\deg\bar C=3, \qquad \gh\bar C=-1.
\eqno\eqn{8}
$$
Thus the BV-quantized action is the sum
$$
S=S_{\rm min}+S_{\rm aux}.
\eqno\eqn{9}
$$

The statement that $S$ satisfies the master equation
\eqn{1} can be also reexpressed in a more traditional
language as a {\it closedness} of $S$ with respect to the
nilpotent BRST operator $s$, where the definition of $s$
is, in a compact notation,
$$
s{\cal A}={\cal F}\equiv d{\cal A}+{\cal A}^2.
\eqno\eqn{10}
$$
Expanding \eqn{10} with respect to \eqn{3}, we explicitly
obtain
$$
sC^a=\half f^{abc}C^bC^c,
$$
$$
sA_\mu^a=(D_\mu C)^a,
$$
$$
sA_{\mu\nu}^{\ast a}=F_{\mu\nu}^a
+f^{abc}A_{\mu\nu}^{\ast b}C^c,
$$
$$
sC_{\mu\nu\lambda}^{\ast a}=D_{[\mu}A_{\nu\lambda]}^{\ast a}
+f^{abc}C_{\mu\nu\lambda}^{\ast b}C^c,
\eqno\eqn{11}
$$
where the first two BRST transformations correspond to the
standard ones. The nilpotency of $s$, $s^2=0$, is equivalent to the generalized
Bianchi identity, ${\cal DF}=0$. Additionally,
$$
s\bar C_{\mu\nu\lambda}^a=B_{\mu\nu\lambda}^a,
$$
$$
s\bar C^{\ast a}=sB_{\mu\nu\lambda}^a=0.
\eqno\eqn{12}
$$

The Landau gauge-fixing condition, $\partial_\mu(\sqrt
gg^{\mu\nu}A_\nu^a)=0$, is introduced to the theory with the
{\it gauge fermion} $\Psi$ of the form
$$
\Psi={1\over6}\int_{{\cal
S}^3}d^3x\,\varepsilon^{\mu\nu\lambda}\Tr\left(g^{\rho\sigma}
A_\rho\partial_\sigma\bar C_{\mu\nu\lambda}\right).
\eqno\eqn{13}
$$
According to the BV prescription each antifield should be
substituted by the (dualized) derivative of $\Psi$ with
respect to the corresponding field. Thus, the partition
function of Chern-Simons theory can be written in the
following path-integral representation
$$
Z_{\rm CS}=\int DA\,DB\,D\bar C\,DC\exp(iS)\equiv \int
d\mu\exp(iS),
\eqno\eqn{14}
$$
where $S$ is given by Eq.~\eqn{9}. In spite of the explicit
dependence of the gauge fermion $\Psi$ on the metric tensor
$g_{\mu\nu}$ the resulting theory is
metric-independent.$^{12}$

\heading{
Observables
}
To encode topology of a link ${\cal L}=\{{\cal C}_i\}$ into a path integral
we shall introduce physical observables in the form of some
auxiliary one-dimensional topological field theory
(topological quantum mechanics) in an external gauge field
$A_\mu$, living on the corresponding loops $\{{\cal C}_i\}$.
The classical action of this theory$^5$ is picked out in
the form of the sum (with respect to $i$) of the terms
$$
S_{{\cal C}_i}(A)=\oint_{{\cal C}_i}dt\,\bar\eta_iD_t^A\eta_i,
\eqno\eqn{1}
$$
where the covariant derivative $D_t^A\equiv d_t+\dot
x_i^\mu(t) A_\mu^a(x(t))t_i^a$, $x_i^\mu(t)$ parametrizes
${\cal C}_i$, and the multiplet of the scalar fields
$\bar\eta_i$, $\eta_i$ is defined in the irreducible
representation $R_i$. The partition function corresponding to \eqn{1}
assumes the following standard form
$$
Z_i(A)=\int D\bar\eta_i\,D\eta_i\exp\left(iS_{{\cal
C}_i}(A)\right).
\eqno\eqn{2}
$$
It can be demonstrated that
our observables are essentially related to the Wilson
ones,$^{13}$ but this fact is not too important for our
further analysis.

We define the topological invariant of the link ${\cal
L}$ as the (normalized) expectation value
$$
\left<\prod_iZ_i(A)\right>
\equiv\left[\int d\mu\exp(iS)\right]^{-1}
\int d\mu\exp(iS)\prod_iZ_i(A).
\eqno\eqn{3}
$$
We can calculate \eqn{3} recursively using the, so-called,
{\it skein relations}. Thus, our present task reduces to the
derivation of the corresponding skein relation. To this
end, we should consider a special link ${\cal L}_{2n}$,
which contains a pair of loops, say ${\cal C}_1$ and ${\cal
C}_2$, where a part of ${\cal C}_1$, forming a small loop
$\ell$ ($\ell=\partial\cal N$), is wrapped round ${\cal C}_2$
$n$-times. In other words, ${\cal C}_2$ pierces $\cal N$ in
$n$ points: ${\cal P}_1,{\cal
P}_2,\ldots,{\cal P}_n$. Such an
arrangement can be interpreted as a preliminary step
towards finding the corresponding monodromy matrix {\bf M}.
Having given the loop $\ell$ we can utilize the Stokes
theorem. Applying the Stokes theorem to \eqn{1} ($i=1$) we
obtain
$$
S_{{\cal C}_1}(A)=S_{{\cal C}_1\setminus\ell}(A)
+\int_{\cal N}d^2\sigma\,\varepsilon^{kl}
\Bigl(D_k^A\bar\eta_1D_l^A\eta_1\qquad
$$
$$
\qquad\left.+\half\partial_kx_1^\mu(\sigma)\partial_lx_1^\nu(\sigma)
F_{\mu\nu}^a(A(x(\sigma)))\bar\eta_1t_1^a\eta_1\right),
\eqno\eqn{4}
$$
where the covariant derivative
$D_k^A\equiv\partial_k+\partial_kx_1^\mu(\sigma)
A_\mu^a(x(\sigma))t_1^a$, and $x_1^\mu(\sigma^1,\sigma^2)$
parametrizes $\cal N$.

In general position, $\cal N$ and ${\cal C}_2$ can
intersect in a finite number of points, and the
contribution to the path-integral coming from these points
can be explicitly calculated. To this end, we should
substitute the curvature in \eqn{4} for the functional
derivative operator
$$
F_{\mu\nu}^a(x)\longrightarrow{4\pi\over
ik}\varepsilon_{\mu\nu\lambda}{\delta\over\delta
A_\lambda^a(x)}.
\eqno\eqn{5}
$$
The substitution \eqn{5} yields an equivalent expression in
\eqn{4} provided the order of the terms in \eqn{3} is such that the
functional derivative can act on $S_{\rm CS}$ producing
$F$.  Essentially, \eqn{5} is a translation operator in a
function space of $A$. Using formal translational
invariance of the product measure $DA$, and functionally
integrating by parts in \eqn{3} with respect to $A$ we
obtain, for each intersection ${\cal P}_m$
($m=1,2,\dots,n$), the {\it monodromy operator}
$$
M=\exp\left[{4\pi\over ik}(\bar\eta_1t_1^a\eta_1)
(\bar\eta_2t_2^a\eta_2)({\cal P}_m)\right].
\eqno\eqn{6}
$$
The expression \eqn{6} is a result of the translation of
$A$ in $S_{{\cal C}_2}(A)$. Strictly speaking the
functional derivative acts on $S$ rather than on $S_{\rm
CS}$, and we should check whether this does not yield some
additional terms. Actually, the substitution \eqn{5}
produces ${\cal F}\vert^0$ rather than $F$, where
``$\vert^0$'' means that only zero ghost-number terms
should be taken. But this difficulty can be easily solved,
because we can substitute $A$ for $\cal A$ in \eqn{1}, and
hence $F$ changes to ${\cal F}\vert^0$ in \eqn{4}. One
should also note that $A$ entering the gauge fermion $\Psi$
(3.13) is as well subjected to the translation, but it is
harmless due to the BV-theorem$^{11}$ on the
$\Psi$-independence of the partition function. Finally, we
can observe that the functional derivative acts trivially, on
geometrical grounds, on the ``kinetic'' term in \eqn{4}.

To calculate the {\it matrix elements} of \eqn{6} one should introduce the
following scalar product
$$
(f,g)={1\over2\pi i}\int
fg\exp(i\bar\eta\eta)d\bar\eta d\eta.
\eqno\eqn{7}
$$
The definition of the scalar product, rather a standard
one, follows from the form of the kinetic term in \eqn{1},
and relates the operator version of the ``evolution'' to the
(holomorphic) path-integral one.
Expanding \eqn{6} in a power series, multiplying with respect to
the scalar product \eqn{7}, and next resumming, we get the {\it
monodromy matrix}
$$
{\bf M}=(\bar\eta_1\bar\eta_2, M\eta_2\eta_1)
=\exp\left({4\pi\over ik}t_1^a\otimes t_2^a\right).
\eqno\eqn{8}
$$
Thus, to the link ${\cal L}_{2n}$, we can associate
the
monodromy matrix
$$
{\bf M}_n={\bf M}^n.
\eqno\eqn{9}
$$

\heading{
Skein Relations
}
The explicit form of the skein relation depends on the
semi-simple Lie group $G$, and on the pair of the
irreducible representations $R_1$, $R_2$. A general method
of the derivation of skein relations from the monodromy
matrix$^{14}$ $\bf M$ bases upon the spectral decomposition of
{\bf M}, and it is given in terms of the Casimir operators
$C_1$, $C_2$, $C_\alpha$ and projectors $P_\alpha$
$$
{\bf M}=\exp\left[{2\pi i\over k}(C_1+C_2)\right]
\sum_\alpha\exp\left(-{2\pi i\over k}C_\alpha\right)P_\alpha,
\eqno\eqn{1}
$$
where $\alpha$ numbers irreducible components in the
Clebsch-Gordan expansion: $R_1\otimes R_2=\bigoplus_\alpha R_\alpha$.

Geometrically, the skein relation consists of a collection
of links ${\cal L}_0,{\cal
L}_2,\dots,{\cal L}_{2N}$, and takes
the form
$$
a_0{\cal L}_0+a_1{\cal L}_2+\cdots+a_N{\cal L}_{2N}=0,
\eqno\eqn{2}
$$
where $N$ is the number of different $C_\alpha$, and all
the $\cal L$'s are identical except a small neighborhood of
$\cal N$. The behavior of ${\cal L}_{2n}$ in the
neighborhood of $\cal N$ has been described in Sect.~4.
The coefficients of the skein relation \eqn{2} are given by
the solution of the corresponding algebraic equation(s)
$$
a_0{\bf M}^0+a_1{\bf M}^1+\cdots+a_N{\bf M}^N=0.
\eqno\eqn{3}
$$
In fact, what we actually obtain is a {\it monodromy} or {\it pure braid} skein
relation (even number of the twists) rather than the {\it braid} one (any
number of the twists). In some cases, we can decompose the
monodromy skein relation into the braid one.

For example, the simplest case of the fundamental representations of
$SU(N)$ group corresponds to the HOMFLY polynomial. The
monodromy skein relation is given by
$$
\exp\left({2\pi i\over kN}\right){\cal L}_0
-2\cos\left({2\pi\over k}\right){\cal L}_2
+\exp\left(-{2\pi i\over kN}\right){\cal L}_4=0,
\eqno\eqn{4}
$$
whereas the standard one is
$$
\exp\left({\pi i\over kN}\right){\cal L}_-
-2\sin\left({\pi\over k}\right){\cal L}_0
-\exp\left(-{\pi i\over kN}\right){\cal L}_+=0,
\eqno\eqn{5}
$$
in accordance with calculations given, for example, in
Ref.~15. The specialization of this group to $SU(2)$
corresponds to the Jones polynomial, whereas other
representations correspond to the Akutsu-Wadati polynomial.
For the fundamental representation of $SO(N)$ group we
obtain the Kauffman-Dubrovnik polynomial.

\heading{
Finishing Remarks
}
There is also a very interesting algebraic side of our
description connected with the notion of {\it
(quasi-)quantum groups}. From that point of view one should
emphasize that the {\it (quasi-)braiding matrices}, implicit in our
construction, satisfy a (quasi-)Yang-Baxter
equation, appearing in the context of quasi-triangular {\it
quasi}-Hopf algebras introduced by Drinfeld,$^{16}$ rather
than the standard one. It confirms some recent
observations,$^{17}$ that it is quasi-quantum group
structure that governs quantum symmetries of some
low-dimensional field theories rather than quantum group
one. Since the quasi-braiding matrix and braiding matrix
are related due to the Drinfeld's theorem both the approaches
should yield equivalent results.

\null

\noindent{\bf
Acknowledgments
}

The author is indebted to Professors E.~Corrigan, H.~D.~Doebner and P.~Goddard
for all arrangements connected with his stay at the Newton Institute.
The author is grateful to Professors R.~Lickorish, J.~Przytycki and C.~Taubes
for interesting discussions.
The work was supported by the Humboldt Foundation.

\null\goodbreak

\noindent{\bf
References
}
\mrm\frenchspacing

\item{1.} D. Birmingham, M. Blau, M. Rakowski and G.
Thompson, {\mit Phys. Rep.} {\mbf 209}, 129 (1991).

\item{2.} E. Witten, {\mit Commun. Math. Phys.} {\mbf 121},
351 (1989).

\item{3.}  L. Smolin, {\mit Mod. Phys. Lett.} {\mbf A 4}, 1091 (1989);
\item{} P. Cotta-Ramusino, E. Guadagnini, M. Martellini
and M. Mintchev, {\mit Nucl. Phys.} {\mbf B330}, 557 (1990).

\item{4.} E. Guadagnini, M. Martellini and M. Mintchev, {\mit Phys. Lett.}
{\mbf B 235}, 275 (1990); {\mit Nucl. Phys.} {\mbf B336}, 581
(1990).

\item{5.} B. Broda, {\mit Mod. Phys. Lett.} {\mbf A 5}, 2747 (1990).

\item{6.} B. Broda, {\mit Phys. Lett.} {\mbf B 262}, 288 (1991).

\item{7.} B. Broda, {\mit Phys. Lett.} {\mbf B 254}, 111 (1991).

\item{8.} M. Blau and G. Thompson, {\mit Ann. Phys.} {\mbf
205}, 130 (1991).

\item{9.} F. Archer and R. M. Williams, {\mit Phys. Lett.}
{\mbf B 273}, 438 (1991).

\item{10.} B. Broda, {\mit Phys. Lett.} {\mbf B 280}, 213 (1992).

\item{11.} I. A. Batalin and G. A. Vilkovisky, {\mit Phys.
Rev.} {\mbf D 28}, 2567 (1983).

\item{12.} M. Blau and G. Thompson, {\mit Phys. Lett.} {\mbf
B 255}, 535 (1991);

\item{ } B. Broda, {\mit Phys. Lett.} {\mbf B 280}, 47 (1992);

\item{} M. Abud and G. Fiore, {\mit Phys. Lett.} {\mbf B
293}, 89 (1992).

\item{13.} B. Broda, {\mit Journ. Math. Phys.} {\mbf 33}, 1511
(1992).

\item{14.} B. Broda, {\mit Phys. Lett.} {\mbf B 271}, 116 (1991).

\item{15.} E. Guadagnini, {\mit Int. Journ. Mod. Phys.} {\mbf
A 7}, 877 (1992).

\item{16.} V. G. Drinfeld, {\mit Alg. Anal.} {\mbf 1}, 114
(1989), in Russian.

\item{17.} G. Mack and V. Schomerus, {\mit Phys. Lett.} {\mbf
B 267}, 213 (1991);

\item{} S. Majid, {\mit Lett. Math. Phys.} {\mbf 22}, 83 (1991).

\bye